\documentclass[11pt]{article}

\usepackage[utf8]{inputenc}
\usepackage[T1]{fontenc}
\usepackage[margin=1in]{geometry}
\usepackage{amsmath,amssymb}
\usepackage{graphicx}
\usepackage{booktabs}
\usepackage{xcolor}
\usepackage{natbib}          
\usepackage{hyperref}
\hypersetup{colorlinks=true, linkcolor=blue, citecolor=blue, urlcolor=blue}
\usepackage{cleveref}

\title{Auditing Belief-Conditioned LLM Agents in\\
Hidden-Information Social Deduction Games}

\author{
Yuan Gao\thanks{Corresponding author: \texttt{yuan.gao.2@student.unimelb.edu.au}} \quad Jiangyi Yang \quad Yao Zhao \quad Yichi Zhang \\[3pt]
University of Melbourne
}
\date{}

\begin{document}
\maketitle

\begin{abstract}
Evaluating LLM agents in hidden-information multi-agent settings is hard: final outcomes are high-variance and rarely reveal why an agent decided as it did. We study this in a 9-player Werewolf environment where agents act under strict, code-level information isolation, and we build an auditable framework that maintains an external belief state over hidden roles, logs belief updates and belief--action deviations as structured evidence, and supports a defensive offline improvement loop that reviews bad cases before any strategy change.

Across 1,080 frozen games spanning belief-disabled, active-belief, kernel-ablation, camp-restricted, consumption-policy, and high-load arms, and including a seed-paired A0/A1 comparison, the active-belief condition is associated with substantially better good-side outcomes: in the 200-seed A0/A1 comparison the good-side win rate rises from 0.205 to 0.390 (paired McNemar $\chi^2=16.4$, $p<0.001$), with fewer irreversible witch-poison errors. We do not, however, attribute this shift to belief content. Direct action--belief consistency is low ($\approx0.21$), and giving belief only to the werewolves helps the good side more than giving it only to the good side, which argues against a simple holder-benefit account; we therefore report the effect as an association and treat its mechanism as unresolved. The contribution is the audit framework itself: it makes the effect measurable, exposes low direct action--belief consistency, rejects an unreliable forced-consumption intervention with evidence, and separates strategy effects from load confounds. We accordingly position external belief in high-noise hidden-information games primarily as an auditable cognitive baseline that also carries decision-relevant signal, turning opaque agent behavior into replayable evidence for safer, controlled iteration.
\end{abstract}

\section{Introduction}
\label{sec:intro}

Large language model (LLM) agents are increasingly used as autonomous decision makers in interactive environments. Many existing evaluations, however, focus on tasks where the full problem state is either directly observable or can be reconstructed from the prompt. Hidden-information multi-agent games pose a different challenge. In these environments, agents must act under partial observability, reason about other agents' private information, interpret potentially deceptive communication, and make decisions whose quality may only become clear after the game ends. Final win rate is therefore a noisy and often insufficient signal: a good decision may lose because of downstream teammate errors, while a poor decision may win because of opponent mistakes.

Social deduction games such as Werewolf provide a compact testbed for this problem. Players have asymmetric private roles, public discussion is cheap and potentially deceptive, and the game outcome depends on a sequence of speech, voting, and role-specific night actions. These properties make the setting useful for studying LLM agents under hidden state, adversarial interaction, and high-variance feedback. At the same time, they expose a core limitation of black-box agent evaluation. If an agent votes for the wrong player, poisons a teammate, withholds a useful claim, or ignores a strong public signal, the final outcome alone does not explain whether the failure came from missing evidence, poor belief tracking, prompt-induced behavior, load-induced LLM failures, or a reasonable strategic override.

This paper studies how to make such agent behavior auditable. We implement a 9-player Werewolf environment in which LLM agents operate under strict information isolation: each agent receives only the public events and private observations available to its role, while the game engine retains the ground-truth role assignment and rule state. The system records structured game events, agent decision traces, belief snapshots, and evaluation reports, enabling post-hoc replay and diagnosis. Rather than treating the LLM as the only carrier of strategy, we separate hidden-state reasoning into an external belief layer that maintains role probabilities over players. This belief layer is not designed as a hard controller. Instead, it provides an auditable cognitive baseline against which LLM actions can be compared.

Our framework has three main components. First, an auditable belief layer updates hidden-role beliefs from structured events. The belief state can be maintained as a shadow signal for evaluation only, or injected into agent contexts as decision support. We also study factorized belief updates that decompose evidence into event direction, source credibility, and base evidence weight, allowing belief changes to be traced back to a small number of interpretable factors. Second, we introduce deviation-guided diagnosis: for key decisions, we compare the action suggested by the belief state with the action actually taken by the LLM agent. Deviations are not automatically errors. They may correspond to harmful mistakes, justified overrides based on language context, or strategic behavior such as deception and misdirection. Third, we use a defensive offline improvement loop. Instead of updating rules or prompts after a single noisy game, the system aggregates batches of games, mines bad cases, generates candidate strategy insights, and leaves versioned updates for human review.

We evaluate this framework using 1,080 clean frozen games spanning belief-disabled agents, active-belief agents, a factorized-versus-additive kernel ablation, camp-restricted belief access, consumption-policy variants, and a high-load replica. The central empirical result is an association: in a 200-seed paired A/B comparison, the active-belief condition raises the good-side win rate from 0.205 to 0.390 (paired McNemar $\chi^2=16.4$, $p<0.001$), and is accompanied by a reduction of irreversible witch-poison errors. We do not, however, attribute this shift specifically to belief content. Direct action--belief consistency is low ($\approx0.21$), camp-restricted arms behave in a way a simple holder-benefit account does not predict (Section~\ref{sec:winrate}), and our belief-disabled baseline retains the belief-oriented prompt structure with the belief content removed rather than a natively designed no-belief prompt. The large gap between the outcome shift and direct consistency leaves the causal pathway unresolved. What the auditable layer does establish is that the effect is measurable and diagnosable: it lets us separate strategy effects from observed infrastructure artifacts such as high-concurrency LLM failures, and reject an unreliable update direction (forced consumption) with evidence.

These findings position external belief in hidden-information LLM agents primarily as an auditable cognitive baseline that also carries decision-relevant signal. When belief is sharp and an agent ignores it the deviation becomes a candidate bad case; when belief is flat, forcing consumption can be counterproductive and should be rejected; and when aggregate belief diagnostics coincide with better outcomes, the system can identify which parts of the pipeline deserve further inspection. The value of the belief layer therefore lies in making agent decisions, and their remaining failures, measurable, replayable, and safer to iterate on under controlled comparison.

This paper makes the following contributions:
\begin{enumerate}
  \item We present an auditable LLM-agent framework for hidden-information social deduction games, with strict information isolation, structured event logging, replayable state, and post-game evaluation.
  \item We introduce an external belief layer for hidden-role reasoning, including shadow and active belief variants, factorized belief updates, and traceable belief snapshots.
  \item We introduce belief--action deviation instrumentation and a conceptual taxonomy (harmful deviation, justified override, strategic deviation) for post-hoc case inspection; large-scale automatic classification and quantitative validation of these deviation types are left to future work.
  \item We demonstrate a defensive offline improvement workflow that mines bad cases from batches of games and supports reviewed, versioned strategy updates rather than single-game online learning.
  \item Through 1,080 clean frozen games, including a 200-seed paired A/B comparison, we show that the active-belief condition is associated with a higher good-side win rate and fewer irreversible low-level errors, while demonstrating---via camp-restricted arms and low direct action--belief consistency---that the causal mechanism is not yet isolated, and that the audit layer nonetheless tracks load confounds and rejects an unreliable forced-consumption intervention with evidence.
\end{enumerate}

Together, these results argue that evaluation infrastructure is a first-class component of hidden-information LLM agents. Before such agents can be safely improved, their cognition must be made observable and contestable. Werewolf is only one domain, but the underlying pattern applies more broadly to adversarial partially observable settings where final outcomes are noisy, decisions require interpretation, and uncontrolled online updates may amplify error rather than learning.

\section{Related Work}
\label{sec:related}

\paragraph{LLM agents in interactive environments.}
Recent work has studied large language models as agents that reason, plan, communicate, and act in interactive environments. ReAct interleaves reasoning traces with environment actions, showing that language models can use intermediate reasoning to guide multi-step decisions \citep{yao2022react}. Generative Agents demonstrates how memory, reflection, and planning can produce believable social behavior in a simulated society \citep{park2023generative}. CAMEL further explores role-playing and communication between LLM agents as a mechanism for cooperative task solving \citep{li2023camel}. Broader benchmarks such as AgentBench and WebArena evaluate LLM agents in long-horizon interactive tasks including web navigation, database use, and tool interaction \citep{liu2023agentbench, zhou2023webarena}. These studies establish LLM agents as an important evaluation object, but they mostly operate in settings where the task state is either public, externally checkable, or recoverable from the interaction history. Hidden-information social deduction games pose a different challenge: agents must reason under asymmetric private information, interpret deceptive language, and make decisions whose quality may not be revealed by the final outcome alone.

\paragraph{Hidden-information and social deduction games.}
Social deduction games such as Werewolf, Mafia, and The Resistance: Avalon provide compact environments for studying deception, persuasion, cooperation, and hidden-role reasoning. Earlier non-LLM work on hidden-role games includes DeepRole, which combines game-theoretic reasoning and learned value estimation for Avalon-style hidden-team play \citep{serrino2019deeprole}. More recent work has used LLMs to construct Werewolf or Mafia agents that speak, vote, deceive, and infer roles from dialogue. Language Agents with Reinforcement Learning for Strategic Play in the Werewolf Game shows that pure LLM agents exhibit systematic decision biases and proposes using an RL policy to select among LLM-generated candidate actions \citep{xu2023werewolfrl}. Werewolf Arena and related benchmarks evaluate LLMs through tournament-style social deduction games, highlighting model differences in deduction, deception, and persuasion \citep{bailis2024werewolfarena}. Newer systems learn strategic language agents through iterative latent-space policy optimization in the Werewolf game \citep{xu2025lspo}, and several recent benchmarks evaluate LLM social deduction from complementary angles: WOLF studies werewolf-based observations of LLM deception and falsehoods \citep{agarwal2025wolf}, MINDGAMES provides a live multi-game arena for social and strategic reasoning in multi-agent LLMs \citep{wang2026mindgames}, and Hidden in Plain Text measures LLM deception quality against human baselines in social deduction games \citep{kao2026hidden}. Our work shares the social deduction setting, but differs in emphasis. Rather than primarily optimizing win rate or dialogue quality, we use Werewolf as a high-noise testbed for auditing hidden-state cognition and diagnosing whether belief-guided interventions are actually useful.

\paragraph{Belief modeling in partially observable multi-agent settings.}
Belief states are a central abstraction in partially observable decision making. Classical POMDP work formalizes belief as a probability distribution over hidden states that can summarize past observations for future decision making \citep{kaelbling1998pomdp}. Interactive POMDPs extend this idea to multi-agent settings by modeling not only the environment state, but also other agents' beliefs, intentions, and policies \citep{doshi2005ipomdp}. This perspective is highly relevant to Werewolf: an agent must infer hidden roles, but also reason about why another player made a claim, cast a vote, or withheld information. Imperfect-information game research has also shown how belief can support planning and policy learning. In Hanabi, Bayesian Action Decoder treats other agents' actions as evidence about hidden information and constructs public belief states for cooperative decision making \citep{foerster2019bad}. Off-Belief Learning studies how to control belief-based reasoning depth and reduce brittle conventions in cooperative partially observable games \citep{hu2021obl}. In competitive imperfect-information games, DeepStack, Deep CFR, and ReBeL combine belief-like public states, search, regret minimization, and learned value functions to handle hidden information in poker-like domains \citep{moravcik2017deepstack, brown2019deepcfr, brown2020rebel}. These methods motivate our use of an explicit belief layer, but our goal is different: we do not attempt equilibrium solving or end-to-end policy optimization. Instead, we engineer belief as an auditable intermediate state for natural-language agents whose decisions remain partly delegated to an LLM.

\paragraph{Opponent modeling and theory of mind.}
Hidden-information social games require models of other agents. Machine Theory of Mind learns to infer agents' latent traits and beliefs from observed behavior \citep{rabinowitz2018tomnet}. Work on agent modeling under partial observability further emphasizes that opponent models should be learned from the controlled agent's available observations, rather than from privileged state information \citep{papoudakis2021agentmodelling}. Neural extensions of I-POMDPs similarly explore how recursive reasoning about other agents can be approximated in learned planners \citep{han2019ipomdpnet}. Recent social deduction systems extend this direction to LLM agents. MultiMind, for example, combines multimodal cues, theory-of-mind reasoning, and planning in a Werewolf-like setting \citep{zhang2025multimind}. These works motivate our role-conditioned and per-agent belief design. However, we deliberately keep our reasoning layer lightweight and auditable: instead of opaque recursive mental-state modeling, we use structured events, role probabilities, belief snapshots, and deviation records that can be inspected after the game.

\paragraph{Evaluation, auditability, and offline improvement.}
A growing body of work argues that LLM agents should be evaluated through trajectories, traces, and failure modes rather than final task success alone. AgentBench and WebArena show that long-horizon agent tasks expose errors in planning, robustness, and instruction following that are not captured by single-step benchmarks \citep{liu2023agentbench, zhou2023webarena}. Reflexion proposes verbal self-improvement through stored reflections over past trials \citep{shinn2023reflexion}. Related agent-improvement systems use generated trajectories, feedback, or skill libraries to improve future behavior. In hidden-information games, however, direct online self-improvement is risky: a single win or loss may be caused by teammate errors, opponent mistakes, hidden information, or infrastructure artifacts rather than the focal decision. Our framework therefore uses a defensive offline improvement loop. It aggregates batches of games, records event logs and decision traces, compares belief-derived recommendations with actual actions, mines bad cases, and leaves candidate strategy updates for review instead of automatically rewriting prompts or rules after one noisy trajectory.

\paragraph{Positioning of this work.}
Our work connects these lines of research: LLM agents in interactive environments, social deduction benchmarks, belief modeling in partially observable games, opponent modeling, and trajectory-level agent evaluation. Prior work shows that belief states can support decision making under hidden information, and that LLM agents can participate in social deduction games. We contribute a complementary perspective: in high-variance hidden-information language games, belief is used primarily as an auditable cognitive baseline that also exhibits decision-relevant signal in our experiments. By recording structured belief updates, belief snapshots, and belief-action deviations under strict information isolation, our system turns opaque LLM decisions into replayable evidence. This enables a more careful account than final win rate alone: the active-belief condition is associated with better good-side outcomes, while the audit layer shows that the mechanism remains unresolved and lets us diagnose flat public evidence, reject an unreliable forced-consumption policy, relate reductions in low-level strategic failures to the outcome shift, and check load-induced confounds against strategy effects.

\section{Problem Setting and System Overview}
\label{sec:setting}

We study LLM agents in a hidden-information social deduction game. The goal of this section is to define the decision setting and the system boundary used throughout the paper. We instantiate the environment as a 9-player Werewolf game, but the core problem is more general: multiple agents act under asymmetric private information, communicate through public language, and make decisions whose quality may only be assessed after hidden state is revealed.

\subsection{Hidden-information social deduction game}
A game consists of a set of players $P=\{1,\ldots,N\}$, hidden role assignments $r_i \in R$, public events, private observations, and a rule-governed state transition process. In our implementation, $N=9$ and the role set contains three werewolves, one seer, one witch, one hunter, and three villagers. Roles determine both information access and action availability. Werewolves know their teammates and nominate night kills; the seer privately checks one player at night; the witch observes the night kill target and may save or poison under role-specific constraints; the hunter may shoot under allowed death conditions; villagers have no private power but participate in public discussion and voting.

The game alternates between night and day phases. At night, role-specific agents take private actions. During the day, agents receive public announcements, speak in sequence, vote, and may enter special resolution phases such as tie discussion, tie revote, exile resolution, last words, or hunter shooting. The game terminates when the rule engine determines a win condition, such as all werewolves being eliminated or werewolves reaching parity with the good side.

This environment creates three difficulties for LLM-agent evaluation. First, agents do not share the same information state. A werewolf, a seer, and a villager may observe the same public dialogue but have different private evidence and strategic incentives. Second, public language is not guaranteed to be truthful. A player may claim a role, accuse another player, hide private information, or deliberately mislead the group. Third, final win rate is a high-variance signal. A locally good decision may be followed by teammate mistakes, and a locally poor decision may still lead to victory because of opponent errors. We therefore treat the game not only as a win-loss benchmark, but as a trajectory-level environment for studying diagnosable agent behavior.

\subsection{Information isolation and agent context}
A central requirement in hidden-information games is that agents must not access privileged state. In our system, the game engine is the only component that stores the ground-truth role assignment and rule state. Agents do not receive the full game state. Instead, before each decision, a context builder constructs a per-agent context containing only the information visible to that agent at that phase. For player $i$ at time $t$, we denote the agent-visible context as
\begin{equation}
x_i^t = f\!\left(H_{\mathrm{public}}^t,\; H_{i,\mathrm{private}}^t,\; S_{\mathrm{visible}}^t,\; A_i^t\right),
\end{equation}
where $H_{\mathrm{public}}^t$ is the public event history, $H_{i,\mathrm{private}}^t$ is the private event history visible to player $i$, $S_{\mathrm{visible}}^t$ contains public player status such as alive/dead state, and $A_i^t$ is the set of legal action types available to that agent at the current phase. The context excludes hidden role maps, truth state, other players' private observations, and internal engine objects.

This separation is implemented as a code-level boundary rather than a prompt-level instruction. Agents are called through serialized data rather than direct references to the game session or stores. The agent-facing context is converted into plain JSON before being passed to the LLM runtime, so that an agent cannot access hidden objects through references. In addition, the system treats role-specific private events as scoped data: werewolf teammate information is visible only to werewolves, seer check results only to the seer, and witch night-kill information only to the witch. Public events such as speeches, votes, deaths, and claims are visible to all alive agents according to the phase rules.

This information-isolation boundary is important for the rest of the paper. Belief states, deviation analysis, and post-game evaluation are only meaningful if the agent could not directly inspect the answer. Any analysis that uses ground-truth roles is therefore performed after the game, not during agent decision making.

\subsection{Agent decision interface}
Each decision is mediated through a structured action interface. Given an agent context $x_i^t$, an LLM agent produces a natural-language or JSON-like response, which is parsed and normalized into a standard action object. The action space is intentionally small and rule-facing:
\begin{equation}
A = \{\texttt{speak},\, \texttt{vote},\, \texttt{night\_kill\_nominate},\, \texttt{check},\, \texttt{save},\, \texttt{poison},\, \texttt{hunter\_shoot},\, \texttt{skip}\}.
\end{equation}
Not every action is available at every phase or to every role. The current legal subset is included in the agent context and enforced by the rule validator. For example, a villager cannot perform a seer check, a seer cannot poison, and a player cannot vote for an invalid target during a tie revote.

The action pipeline is designed to separate open-ended LLM generation from rule-governed state transitions. After an LLM response is produced, an action parser extracts the intended structured action. A canonicalization layer then removes or rejects non-standard action types, role leakage, meta-AI phrases, or chain-of-thought-like text that should not enter the game state. The rule validator checks whether the action is legal under the current phase and role. If parsing or validation fails, a fallback policy selects a safe legal action when possible, allowing the game to continue while recording the failure. Only validated actions are applied by the engine.

\begin{figure}[t]
\centering
\includegraphics[width=\linewidth]{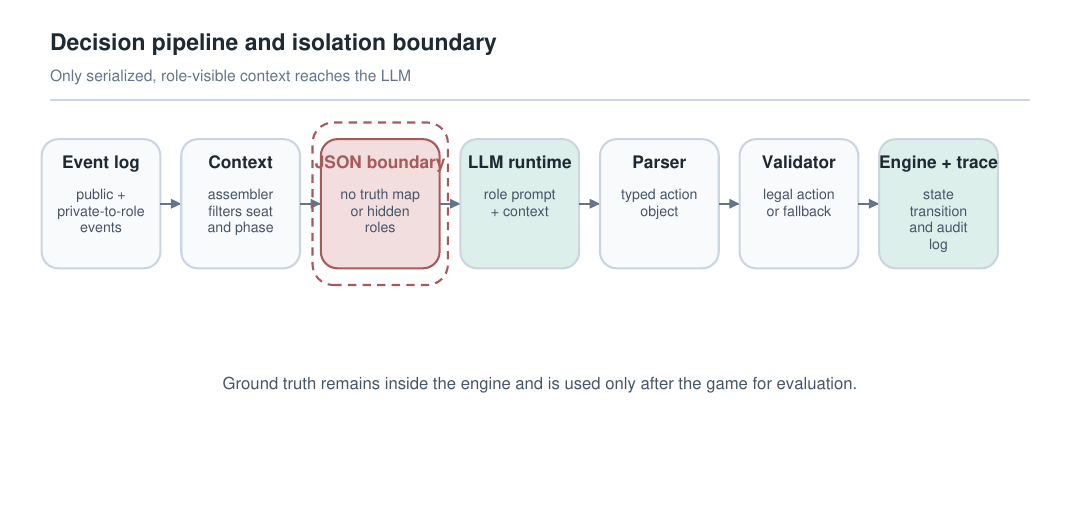}
\caption{Decision pipeline and information-isolation boundary. The per-agent context is serialized before the LLM runtime; parser, canonicalizer, rule validator, and fallback sit between LLM output and the engine.}
\label{fig:pipeline}
\end{figure}

This interface serves two purposes. It makes the game engine robust to malformed LLM outputs, and it ensures that all decisions can be logged in a comparable form. As a result, speeches, votes, night actions, invalid outputs, retries, canonicalization events, and fallbacks can all be linked to the same decision trace (see \Cref{fig:pipeline}).

\subsection{Replayable event log and evaluation interface}
The system records each game as a structured trajectory. The event log contains public and private game events, including phase starts, speeches, votes, night actions, deaths, role-specific private observations, and game-over events. Separately, decision traces record the agent id, role, phase, prompt or agent version, input summary, parsed action, output quality flags, retry counts, fallback status, and token or latency metadata when available. Belief snapshots and belief update records are stored as additional observability data when belief tracking is enabled.

This trajectory representation allows the game to be replayed from logs and analyzed after completion. During runtime, agents only receive their own visible contexts. During post-game evaluation, the evaluator can access the final truth state to compute outcome-dependent metrics such as whether a vote hit a werewolf, whether the witch poisoned a good-side player, or whether a belief recommendation pointed to the correct camp. This separation lets us compare agent behavior against ground truth without leaking ground truth into the original decision process.

The evaluation objective is therefore broader than final win rate. We are interested in whether a system can answer questions such as: Did the agent receive only legal information? Did the LLM output require parsing repair or fallback? Did a belief state identify the correct suspect before a vote? Did the agent follow or ignore the belief recommendation? If it ignored the recommendation, was the deviation harmful, justified, or strategically useful? Did a proposed strategy update actually improve behavior, or did it amplify noise? These questions require structured replay and post-hoc auditability, not only game outcomes.

\subsection{Overview of the framework}
\begin{figure}[t]
\centering
\includegraphics[width=\linewidth]{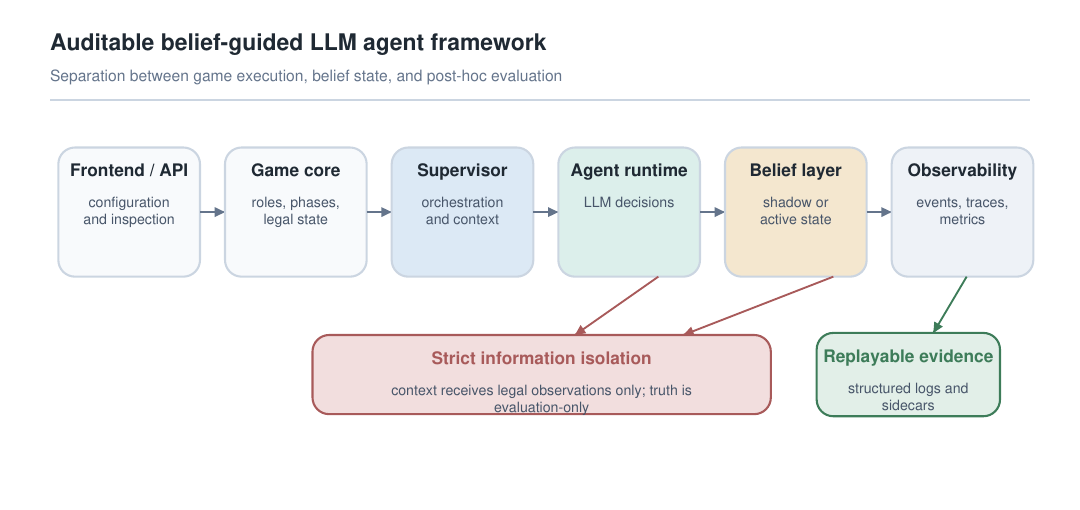}
\caption{Overall framework: layers from frontend/API through game core, supervisor/context, agent runtime, belief maintenance, and observability/evaluation.}
\label{fig:framework}
\end{figure}

At a high level, each phase follows a fixed decision loop. The engine determines the current phase and required actors. For each required actor, the context builder constructs an information-isolated agent context. The LLM runtime generates an action, which is parsed, canonicalized, validated, and, if necessary, replaced by a safe fallback. The engine applies validated actions and emits structured events. The event log, trace store, and belief store then record the resulting state for replay and evaluation (\Cref{fig:framework}).

This architecture supports the method introduced in the next section. The belief layer can be maintained in the background as a shadow diagnostic signal or injected into the agent context as decision support. Deviation analysis can compare belief-derived recommendations with the agent's actual actions because both belief snapshots and actions are logged. The offline improvement loop can mine batches of completed games because the system records comparable trajectories across agent versions and experimental arms. In this way, the system turns a hidden-information language game into an auditable experimental platform for studying LLM-agent cognition under partial observability.

\section{Method}
\label{sec:method}

This section describes the auditable belief-guided framework used in our Werewolf environment. The framework is designed around a conservative principle: belief should make hidden-state reasoning observable and testable, but should not be treated as an unquestionable controller of LLM behavior. We therefore separate the framework into five components: an auditable belief layer, a factorized update mechanism, belief consumption policies, deviation-guided diagnosis, and a defensive offline improvement loop.

\subsection{Auditable Belief Layer}
For each agent, the system may maintain a belief state over the possible roles of other players. Let $b_i^t(j)$ denote agent $i$'s belief about player $j$'s hidden role at time $t$. In implementation, this is represented as a role distribution rather than a single scalar suspicion score. For example, a target player may simultaneously have nonzero probability of being a werewolf, seer, witch, hunter, or villager. This is important because different roles require different interpretations: the same public claim may increase both the probability that a player is the true seer and the probability that the player is a werewolf bluffing as seer.

The belief layer is updated from structured game events rather than directly from the full raw transcript. Events include public claims, check results, votes, deaths, hunter shots, and role-specific private observations. Each update produces a new belief snapshot and can be associated with the event that triggered it. This makes belief a replayable intermediate state: after the game, we can inspect not only what the belief was, but also why it changed.

We distinguish two modes of belief usage. In the shadow mode, belief is maintained in the background for evaluation and diagnosis, but is not shown to the LLM agent. This allows us to compare belief-free decisions with a belief-derived baseline without changing the agent's runtime context. In the active mode, selected belief summaries, such as top suspected werewolves or role likelihoods, are injected into the agent context as decision support. In both modes, belief is not treated as ground truth. It is a subjective state derived only from information visible to the observer.

This distinction is crucial for information safety. The belief updater used during the game does not read the hidden role map as an input for real-time updates. Ground-truth roles are used only after the game for evaluation, such as computing whether a belief recommendation or vote target was correct.

\subsection{Factorized Belief Update}
\label{sec:factorized}
A direct rule table for hidden-role belief updates would be difficult to maintain. The effect of an event depends on the event type, the speaker, the observer, the target, the current phase, and the observer's prior beliefs. Enumerating all combinations would quickly become brittle and opaque. We therefore use a factorized update view. For a belief update affecting observer $i$'s belief about target $j$ and role $r$, we write the update abstractly as
\begin{equation}
\Delta_{i,j,r} = w(e)\cdot d(e,r)\cdot c_i(s_e),
\end{equation}
where $e$ is the triggering event, $w(e)$ is a base evidence weight for the event type, $d(e,r)$ is the direction of evidence for role $r$, and $c_i(s_e)$ is observer $i$'s credibility estimate for the source $s_e$ of the event. For instance, a public seer claim has different implications depending on whether the observer currently regards the speaker as credible. A trusted seer claim should have stronger influence than the same claim from a player already considered suspicious.

\begin{figure}[t]
\centering
\includegraphics[width=\linewidth]{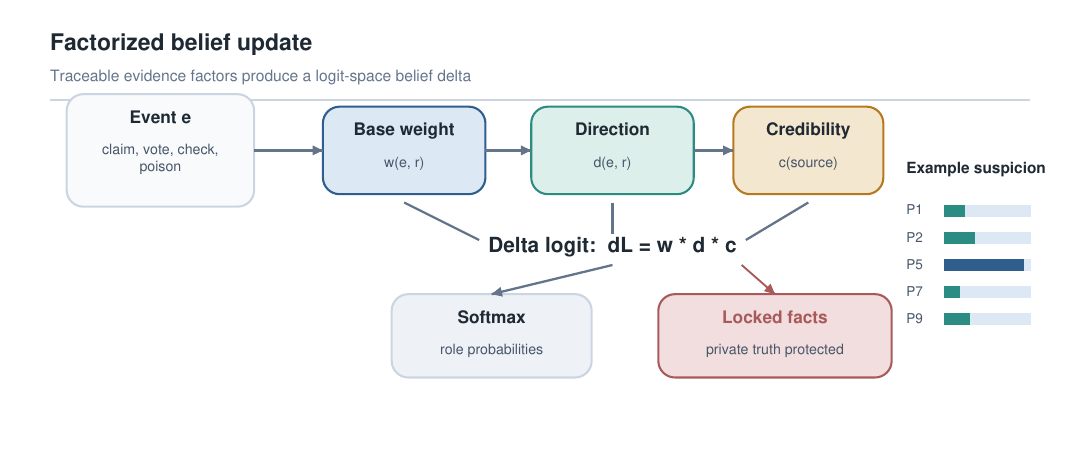}
\caption{Factorized belief update. Each event contributes a logit-space delta decomposed into base weight, evidence direction, and source credibility; locked beliefs are protected from ordinary public-event updates.}
\label{fig:belief}
\end{figure}

In the simpler additive variant, the system applies bounded probability deltas followed by normalization. This version is easy to inspect and was useful as an initial engineering baseline. In the factorized variant, updates are applied in log-odds space:
\begin{equation}
\mathrm{logit}\!\left(b_i^{t+1}(j,r)\right) = \mathrm{logit}\!\left(b_i^{t}(j,r)\right) + \Delta_{i,j,r}.
\end{equation}
Log-odds updates have two practical advantages. First, repeated evidence combines additively in logit space while the resulting probability remains within $[0,1]$. Second, update strength can be controlled through a small number of interpretable factors rather than a large table of special cases (\Cref{fig:belief}). We emphasize that this is a Bayesian-inspired, rule-weighted heuristic rather than an exact Bayesian posterior estimator: the factors $w$, $d$, and $c$ are hand-specified rather than learned, and the update carries no optimality or calibration guarantee. Its purpose is to make each belief change traceable to a small set of named factors, not to solve inference optimally. The concrete factor values, clamp ranges, normalization order, locked-state conditions, and duplicate-evidence handling will be included in the accompanying public release (Code and Data Availability).

The belief state is recursive. The current belief snapshot summarizes previous evidence and becomes the prior for the next event. The update mechanism does not need to replay the full game history at every step. This keeps belief updates efficient and makes each belief transition attributable to a local event. Some role beliefs may also be locked when the observer has private certainty, such as a werewolf knowing a teammate. Locked beliefs are protected from ordinary public-event updates.

\subsection{Belief Consumption Policies}
The belief layer can influence agents in different ways. We study belief consumption as an explicit policy rather than assuming that injecting belief into the prompt is always beneficial.

The first policy is \emph{no consumption}. In this setting, agents do not receive belief information, while the system may still maintain shadow belief for post-hoc analysis. This corresponds to the belief-disabled condition (A0 in our experiments). The second policy is \emph{active belief injection}. Here, the agent context includes belief summaries such as the top suspected players or role likelihoods. The prompt instructs the agent to treat belief as a subjective signal, not as truth. The LLM remains responsible for producing the final action and public language. The third policy is \emph{confidence-gated consumption}. This policy only encourages or requires alignment with belief when the belief signal is sufficiently sharp. If the belief distribution is flat, the agent should not be forced to follow it. This is important in Werewolf because public evidence is often weak. A weak belief recommendation can be worse than no recommendation if it creates false confidence.

We therefore distinguish between belief availability and belief authority. A belief may be available to the agent, but it does not automatically override role strategy, private information, legal action constraints, or natural-language context. This design allows belief to support decision making while still preserving the LLM's ability to make strategic deviations.

\subsection{Deviation-Guided Diagnosis}
Because belief is treated as a baseline rather than a controller, disagreement between belief and action is informative. We define a deviation when the action suggested by belief differs from the action chosen by the LLM agent at a key decision point. For example, if belief ranks player $P_3$ as the most suspicious werewolf candidate but the agent votes for $P_5$, the system records a belief-action deviation.

\begin{figure}[t]
\centering
\includegraphics[width=\linewidth]{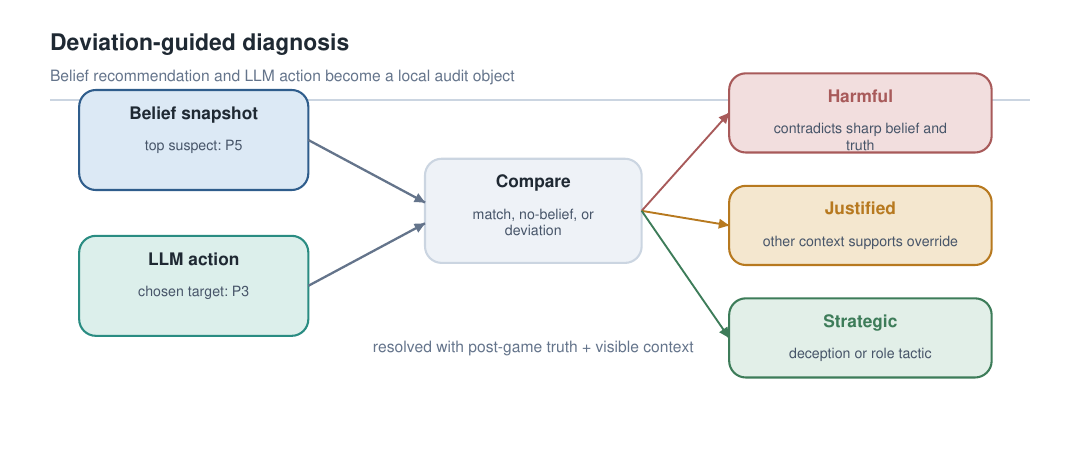}
\caption{Deviation-guided diagnosis. The framework compares belief-recommended targets with actual LLM actions and uses post-game truth plus context to inspect deviations.}
\label{fig:deviation}
\end{figure}

A deviation is not automatically an error. We classify deviations conceptually into three types. A \emph{harmful deviation} occurs when the agent ignores a useful belief signal and the decision harms its side, such as a good-side player voting away from a strong wolf suspect without a clear reason. A \emph{justified override} occurs when the belief signal is incomplete or flat, and the LLM uses other visible context to make a better decision. A \emph{strategic deviation} occurs when the agent intentionally acts against the belief baseline for game-theoretic reasons, such as a werewolf voting against a teammate to reduce suspicion (\Cref{fig:deviation}).

The key point is that deviation makes LLM behavior auditable. Without belief, a vote is just an action. With belief, the same vote can be interpreted relative to a visible cognitive baseline. During post-game evaluation, ground truth can be used to assess whether the deviation was helpful, harmful, or neutral, but that truth is not available to the agent when it acts. This mechanism also helps avoid over-trusting the belief system. If belief is flat and the agent deviates, the deviation may not be meaningful. If belief is sharp and repeatedly ignored in harmful ways, the case becomes a candidate for prompt or strategy review. Thus, deviation analysis evaluates both the agent and the belief layer.

\subsection{Defensive Offline Improvement Loop}
\begin{figure}[t]
\centering
\includegraphics[width=\linewidth]{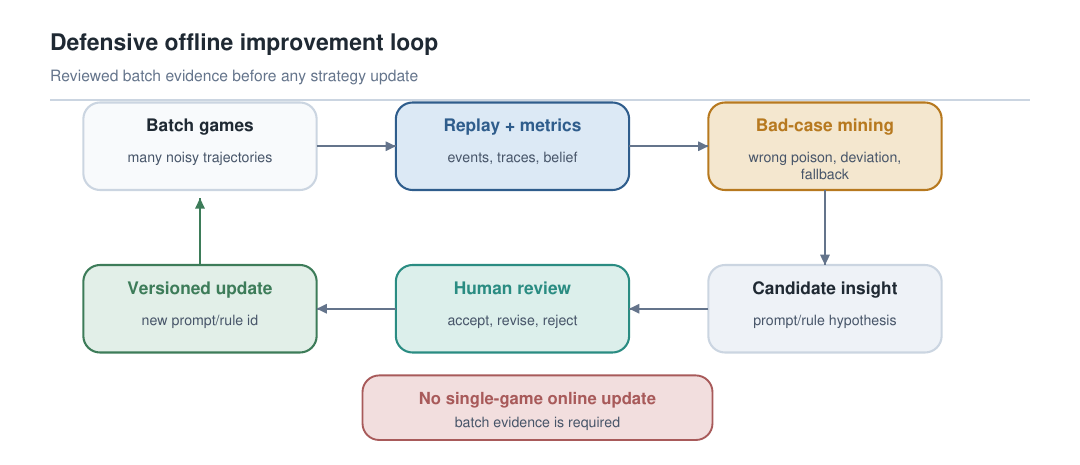}
\caption{Defensive offline improvement loop. Strategy changes are proposed from batch evidence and reviewed before any versioned prompt or rule update.}
\label{fig:loop}
\end{figure}

Hidden-information games are high-noise environments. Updating an agent after a single game is risky because the final outcome may be caused by many factors unrelated to the focal decision. A team may win despite poor reasoning, or lose despite a locally correct action. Infrastructure artifacts such as LLM timeouts or fallback behavior can also distort apparent strategy quality. For this reason, our framework uses a defensive offline improvement loop (\Cref{fig:loop}). The loop has five steps:
\begin{enumerate}
  \item Run a batch of games under fixed experimental conditions.
  \item Store event logs, decision traces, belief snapshots, and final truth states.
  \item Compute aggregate metrics and mine bad cases such as harmful deviations, wrong poison actions, inconsistent votes, invalid actions, or load-induced failures.
  \item Generate candidate strategy insights or belief-rule adjustments from the batch.
  \item Review and version any accepted changes before running another batch.
\end{enumerate}

The loop is deliberately not fully automatic. Candidate updates are treated as hypotheses, not as guaranteed improvements. A suggested prompt change, belief-rule adjustment, or strategy snippet should be validated against additional games before becoming part of a new experimental arm. This design prevents single-game noise from directly modifying the agent policy.

The same loop also supports negative findings. If an intervention fails, the system can reject it with evidence. For example, forcing agents to follow weak belief recommendations may reduce alignment or fail to improve decisions when the underlying belief distribution is flat. In that case, the correct update is not to force stronger consumption, but to diagnose why belief lacks separability at the decision point.

Together, these components make the framework auditable rather than merely belief-guided. The belief layer supplies an interpretable hidden-state baseline; consumption policies control how that baseline enters the agent context; deviation analysis turns agent choices into diagnosable evidence; and offline review converts batches of noisy games into safer, versioned improvement candidates.

\section{Experimental Setup}
\label{sec:setup}

We evaluate the framework in a 9-player Werewolf environment. The experiments are designed to answer three questions: (1) whether the system can run hidden-information LLM-agent games safely and reproducibly; (2) whether active-belief conditions are associated with changes in game outcomes or decision quality; and (3) whether belief traces and deviations provide useful diagnostic signals when the outcome effect is not fully explained.

\subsection{Environment and game configuration}
All experiments use the same 9-player Werewolf rule configuration: three werewolves, one seer, one witch, one hunter, and three villagers. Each game proceeds through role assignment, night actions, day announcements, sequential discussion, voting, tie resolution when needed, exile resolution, hunter shooting when legally triggered, win checking, and game termination. The game engine is responsible for the ground-truth state, legal action validation, phase transitions, deaths, exile, hunter-shot resolution, and win conditions. Agents act through the fixed eight-action interface defined in \Cref{sec:setting}. The legal subset depends on the current role and phase; invalid or malformed LLM outputs are parsed, canonicalized, validated, and, if necessary, replaced by a safe legal fallback, and all such events are logged.

Information isolation is enforced throughout the game. Before each agent decision, the context builder constructs an agent-specific context containing public events, role-visible private events, alive/dead status, legal actions, and optional belief summaries depending on the experimental arm. The runtime agent does not receive the ground-truth role map, hidden roles of other players, other agents' private observations, or direct references to game-engine state. Ground truth is used only after the game for evaluation.

\subsection{Compared agent arms}
\label{sec:arms}
We compare several agent variants that differ in whether and how belief is maintained and consumed.

\textbf{Belief-disabled ablation (A0).} The belief-disabled arm receives no belief information in the agent context. The agent acts only from public events, private role observations, role prompts, and legal action constraints. The system may still maintain shadow belief in the background for post-hoc diagnosis, but this shadow belief is not visible to the agent. Importantly, A0 preserves the broader belief-oriented prompt structure and removes only the belief content; it is therefore a belief-content ablation rather than a natively designed no-belief prompt, and it does not by itself isolate belief content from prompt-structure effects. A native no-belief baseline (denoted A0-native) is left to future work (Section~\ref{sec:futurework}) and is not part of the results reported here.

\textbf{Active belief guidance.} The belief-guided arm injects belief summaries into the agent context, such as top suspected werewolf candidates or role likelihoods. The prompt instructs the agent to treat belief as a subjective signal rather than truth. The LLM still produces the final action, and the rule engine still validates legality.

\textbf{Factorized belief variant.} The factorized variant uses the belief update mechanism described in \Cref{sec:factorized}, where event direction, source credibility, and base evidence weights jointly determine belief updates. Compared with the additive belief baseline, this variant is intended to make belief updates more interpretable and less dependent on hand-enumerated special cases.

\textbf{Mixed-access belief variants.} To separate camp-specific effects, we also evaluate mixed variants in which only one side receives belief information. In the villagers-only variant, good-side agents can consume belief while werewolves do not. In the wolves-only variant, werewolves can consume belief while good-side agents do not. These arms help distinguish whether belief is useful because it improves hidden-role inference, because agents actually consume it, or because one camp can operationalize the signal better than the other.

\textbf{Consumption-policy variants.} For controlled studies, we evaluate prompt variants that encourage or require stronger alignment with belief recommendations. These variants test whether increasing belief consumption improves decisions, or whether forcing agents to follow weak belief signals can be harmful when the belief distribution is flat.

Each arm is identified by its game configuration, prompt version, belief kernel, model provider, random seed range, and concurrency setting. This is important because comparing arms across different model versions or load regimes can create misleading conclusions.

\subsection{Metrics}
We report both outcome metrics and diagnostic metrics. The primary outcome metric is win rate by camp. However, because Werewolf outcomes are high variance, win rate is not treated as the only success criterion. We additionally report role-conditioned and decision-conditioned metrics where appropriate.

For belief quality, we compute the \emph{top-1 hit rate} as the fraction of relevant decision points where the highest-ranked werewolf suspect is actually a werewolf after ground truth is revealed. We also compute top-2 hit rate when useful, because a belief state may identify the correct suspect set without ranking the exact top player correctly. We measure belief separability as the difference between average werewolf belief assigned to true werewolves and average werewolf belief assigned to non-werewolves at key decision points. Higher separation indicates that the belief state distinguishes wolves from good-side players more clearly. We also report entropy or flatness of the belief distribution, since a high hit rate with very low margin may still be too weak to guide decisions safely.

For belief consumption, we measure whether the agent's action aligns with the belief recommendation at relevant decisions, especially day votes and high-risk role actions. This produces an adoption or consistency rate. We also record deviations, where the agent chooses a target different from the belief-recommended target. For decision quality, we track selected low-level errors and high-risk actions, including wrong poison actions, incorrect hunter shots, votes against the belief recommendation, illegal actions, fallback usage, parsing failures, LLM errors, retry counts, and canonicalization events. These metrics help distinguish strategy failures from infrastructure failures. Finally, we track resource and robustness metrics such as token usage, context truncation, context-budget exceedance, latency, and load-related LLM failures, because infrastructure artifacts can bias apparent strategic performance.

\subsection{Evaluation protocol}
Each game is logged as a structured trajectory containing game events, agent decision traces, belief snapshots, and post-game truth. We evaluate agents only from data generated under the same action interface and rule engine. Ground-truth roles are never used during runtime belief updates or agent decisions; they are used only in post-game evaluation.

To reduce confounding, comparisons should be run under fixed or recorded model versions, seed ranges, concurrency limits, prompt versions, and belief settings. In particular, we treat concurrency as an experimental variable rather than an implementation detail. High-concurrency runs can increase LLM timeout or failure rates, and fallback behavior may affect the two camps asymmetrically. Therefore, any arm with substantially different LLM error or fallback rates is analyzed separately rather than treated as a clean strategy comparison. For large-scale diagnosis, we use batch-level aggregation. A batch report summarizes total games, completed games, failed games, winner distribution, decision quality statistics, belief quality statistics, and role/action-specific error rates. For case studies, we inspect individual trajectories using the event log, decision trace, belief curve, and final truth state.

Our main evaluation corpus is a single frozen experiment of 1,080 games run under a fixed configuration (one model, \texttt{deepseek-chat}; temperature 0.6; contract version 2.2; recorded git commit), with paired seeds used for the main A/B comparisons. It comprises eight arms: a belief-disabled ablation (A0, 200 games) and an active-belief arm (A1, 200 games) sharing seeds 7000--7199, a villagers-only and a wolves-only belief arm (150 games each), an additive-versus-factorized kernel ablation (120 games), a seed-paired consumption-policy A/B (80 games per arm), and a high-concurrency replica (100 games). All arms completed with zero failed games and a 0\% LLM-error rate. When reporting results, we separate large-batch aggregate findings (A0/A1 at $n=200$ per arm) from smaller controlled A/B studies (such as the $n=80$ consumption-policy arms), and we avoid interpreting small-sample win-rate differences as statistically conclusive.

\subsection{Reproducibility controls}
For each experiment, we record the game id, seed or seed range when available, model name, prompt version id, belief mode, belief kernel, whether belief is injected into each agent, concurrency setting, and output directories for events, traces, belief states, and batch reports. This metadata is necessary for reproducing aggregate metrics and for separating true strategy effects from run-configuration artifacts.

We also use unit and integration tests to verify core invariants: contract schemas remain valid, agents receive information-isolated contexts, the rule engine terminates games without phase-stuck loops, invalid actions do not pass through to the engine, shadow belief is not injected into belief-disabled agents, and belief updates do not read ground-truth roles during runtime. These tests do not replace empirical evaluation, but they establish that the experimental substrate enforces the hidden-information assumptions required by the paper.

\section{Results and Analysis}
\label{sec:results}

This section reports the main empirical findings. We focus on three questions: whether the system provides a stable hidden-information substrate, whether active-belief conditions are associated with changes in game outcomes, and what diagnostic value belief provides for characterizing the outcome shift and remaining failure modes.

\subsection{System stability and information safety}
The frozen 1{,}080-game corpus shows that the platform runs complete 9-player LLM-agent games at scale while preserving the hidden-information boundary. In the 200-game active-belief arm (A1), across 11{,}724 recorded agent decisions the LLM-error and retry rates were both 0\%, rule-validation fallback fired on only 28 decisions (0.24\%), and no role-leakage, meta-AI leakage, or chain-of-thought leakage passed through the action canonicalization layer. No context exceeded the configured 4000-token budget, and the belief and replay infrastructure recorded 82{,}967 belief saves and 746{,}703 belief-curve points for that arm alone (\Cref{tab:stability}). These results support the use of the environment as an experimental substrate: agents complete games, hidden information remains isolated, and every trajectory is structured enough for post-game analysis.

We also verified the implementation through unit and integration tests: the frozen run metadata records 1{,}138 passing tests (0 skipped), covering contracts, rule validation, information isolation, action parsing and canonicalization, game termination, belief updates, replay, stores, batch runners, and API-level services. (An earlier 10-game real-LLM smoke test used during development---483 decisions, 482 completed and one recovered by retry, with zero leakage through canonicalization---is an engineering check only and is not part of the main corpus or its statistics.) These tests do not prove strategic quality, but they establish that the experimental substrate enforces the assumptions required for hidden-information evaluation. \Cref{tab:stability} summarizes the stability and safety metrics.

\begin{table}[t]
\centering
\caption{Stability and safety metrics from the frozen package and A1 clean-batch runtime statistics. The information-safety row counts leakage that \emph{passed through} canonicalization; separately, one CoT attempt was intercepted and sanitized by the canonicalizer (0 passed through).}
\label{tab:stability}
\small
\begin{tabular}{lll}
\toprule
Dimension & Metric & Value \\
\midrule
Test suite & passed / skipped tests & 1,138 / 0 \\
Decision scale & total decisions (A1, 200 games) & 11,724 \\
LLM reliability & llm\_error / retry & 0 / 0 \\
Rule fallback & fallback\_used & 28 (0.24\%) \\
Information safety & role/meta/CoT leakage through canonicalization & 0 / 0 / 0 \\
Context budget & context-budget exceedances & 0 \\
Observability & belief saves / curve points (A1) & 82,967 / 746,703 \\
\bottomrule
\end{tabular}
\end{table}

\subsection{The active-belief condition is associated with a higher good-side win rate under paired evaluation}
\label{sec:winrate}
The main large-batch result is a clear outcome shift. We compare a belief-disabled arm (A0) and an active-belief arm (A1) that share the same model, game configuration, prompt family, and random seed range (7000--7199, $n=200$ each), with the controlled difference being whether factorized belief is injected into the agent context (A0 keeps the belief-oriented prompt structure with the belief content removed; see Section~\ref{sec:arms}). Under this paired design, the good-side win rate rises from 0.205 in A0 to 0.390 in A1. A paired McNemar test over the 200 shared seeds is significant (58 seeds flip to a good-side win under belief versus 21 the other way; $\chi^2=16.4$, $p<0.001$), and the 95\% confidence intervals ($\pm5.6$pp and $\pm6.8$pp) do not overlap.

This effect is not explained by the load or infrastructure metrics we record. Both arms recorded a 0\% LLM-error rate and no context-budget violations, so the difference is not driven by asymmetric LLM failures under load. Because the arms are seed-paired and share the same model, contract version, and prompt family, the comparison supports treating active belief as the primary experimental difference rather than model drift or concurrency.

\begin{figure}[t]
\centering
\includegraphics[width=\linewidth]{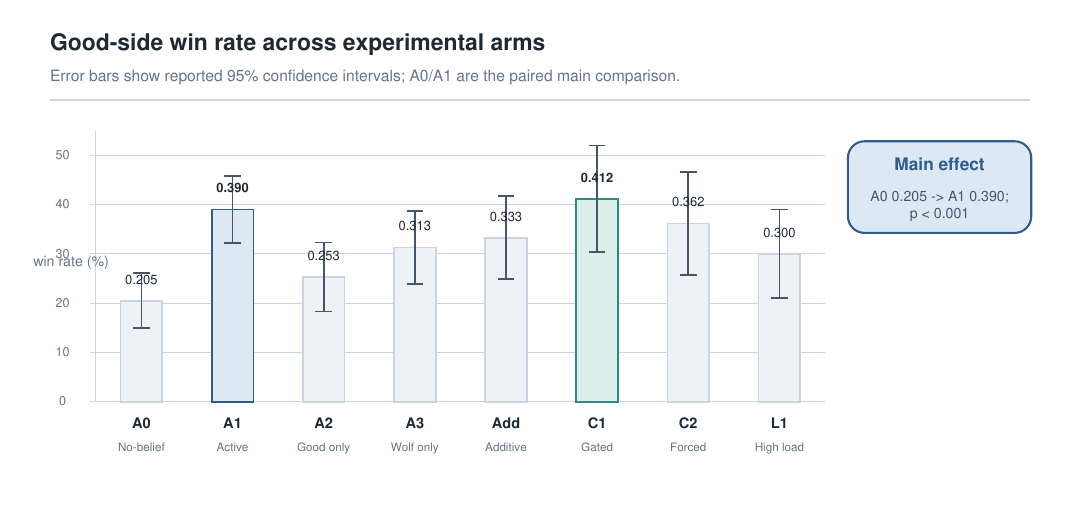}
\caption{Good-side win rate across experimental arms. Error bars show the reported 95\% confidence intervals; A0/A1 is the main seed-paired comparison, while the smaller controlled arms are used as supporting diagnostics rather than standalone claims.}
\label{fig:winrate}
\end{figure}

The camp-restricted arms complicate any simple reading of this shift. Giving belief only to the good side (villagers-only, 0.253) raises the good-side win rate \emph{less} than giving it only to the werewolves (wolves-only, 0.313), and both are below giving it to both camps (0.390). This pattern contradicts a simple holder-benefit explanation in which belief primarily improves the hidden-role inference of the agents that receive it: if it did, villagers-only belief should help the good side most, not least. We therefore treat the mechanism as unresolved rather than interpreting the camp-restricted arms as evidence of improved holder-side inference; the effect may instead operate through changes in how the belief section reshapes agent behavior, or through asymmetric interactions between the two camps. We report these smaller arms as exploratory diagnostics, not as standalone causal claims, and we do not apply multiple-comparison correction across them. The kernel ablation is likewise exploratory: the factorized variant is cleaner and more interpretable than the additive baseline (good-side win rate 0.390 versus 0.333), but this non-paired difference does not by itself establish that one kernel is superior. \Cref{fig:winrate} visualizes the arm-level outcome pattern, and \Cref{tab:winrate} reports win rate and aggregate belief diagnostics across arms.

\begin{table}[t]
\centering
\caption{Good-side win rate and aggregate belief diagnostics. Belief diagnostics are sidecar aggregates over belief-targeted decisions and are not strict good-side vote-only accuracy estimates.}
\label{tab:winrate}
\small
\resizebox{\linewidth}{!}{%
\begin{tabular}{llrrrrrrr}
\toprule
Arm & Description & $n$ & Good WR & 95\% CI & Top-1 & Top-2 & Separation & Consistency \\
\midrule
A0 & Belief-disabled ablation & 200 & 0.205 & $\pm$5.6pp & -- & -- & -- & -- \\
A1 & Active belief, factorized & 200 & \textbf{0.390} & $\pm$6.8pp & 0.686 & 0.825 & 0.356 & 0.207 \\
A2 & Good-side belief only & 150 & 0.253 & $\pm$7.0pp & 0.493 & 0.773 & 0.043 & 0.370 \\
A3 & Wolf-side belief only & 150 & 0.313 & $\pm$7.4pp & 0.899 & 0.899 & 0.803 & 0.092 \\
ABL-add & Additive belief kernel & 120 & 0.333 & $\pm$8.4pp & 0.668 & 0.804 & 0.347 & 0.201 \\
C1 & Gated consumption baseline & 80 & 0.412 & $\pm$10.8pp & 0.653 & 0.802 & 0.343 & 0.250 \\
C2 & Stronger consumption prompt & 80 & 0.362 & $\pm$10.5pp & 0.688 & 0.840 & 0.352 & 0.280 \\
L1 & High-concurrency A1 replica & 100 & 0.300 & $\pm$9.0pp & 0.706 & 0.825 & 0.359 & 0.284 \\
\bottomrule
\end{tabular}}
\end{table}

\subsection{Aggregate belief diagnostics indicate usable signal, while direct action--belief consistency remains low}
\label{sec:diagnostics}
Aggregate belief diagnostics indicate that the active-belief arm has usable hidden-role signal. Across belief-targeted decisions recorded in the sidecar, the top-ranked werewolf suspect is a true werewolf 68.6\% of the time (top-1 hit rate), and a true werewolf falls within the top two suspects 82.5\% of the time (top-2 hit rate). Average wolf-vs-good separability is 0.356 and the average Brier score is 0.160, indicating that belief distinguishes wolves from good-side players in a usable, if not sharp, way. These aggregate diagnostics mix action types and agent camps: werewolves trivially know their teammates, which inflates the pooled hit rate, so the good-side-only rate at vote decisions is substantially lower (closer to the villagers-only arm) and only modestly above the roughly 0.375 chance level of picking one of three werewolves among the remaining players. We therefore treat these numbers as evidence of available, usable signal rather than as a strict good-side vote-only accuracy estimate. Together with the paired outcome gain and the wrong-poison reduction, they are consistent with belief contributing useful decision support.

\begin{figure}[t]
\centering
\includegraphics[width=0.9\linewidth]{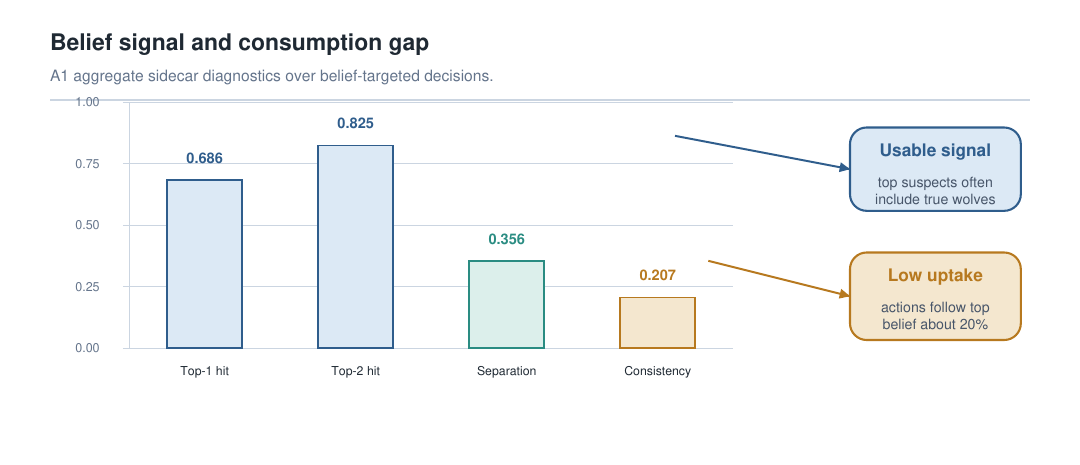}
\caption{Belief signal and consumption gap in the A1 active-belief arm. Aggregate belief diagnostics show usable hidden-role signal, while action--belief consistency remains low.}
\label{fig:signal-consumption}
\end{figure}

At the same time, direct belief consumption is low: the action--belief consistency rate --- how often the agent's action exactly matches the top belief recommendation --- is only about 0.21 in the active-belief arm (\Cref{fig:signal-consumption}). This measures only exact top-target agreement; it does not capture candidate-set narrowing, rank-order shifts, changes in speech caution, or indirect linguistic and multi-agent effects, so it is a lower bound on how much belief is used. The large gap between the outcome shift (Section~\ref{sec:winrate}) and this low direct-consistency figure leaves the causal pathway unresolved: belief may influence behavior without producing exact top-target agreement, and prompt structure or other behavioral changes may also contribute. We therefore do not claim that ``insufficient consumption'' is \emph{the} bottleneck; we report the gap itself as an open question and, in Section~\ref{sec:forced}, test one intervention against it.

Separability, though usable, remains moderate: the top suspect often only weakly differs from the next candidates, and distributions are still somewhat flat. This matters for consumption-policy design. When the margin is small, blindly following the top recommendation is not always justified --- a cautious LLM may rationally weigh other context. A good consumption policy must therefore be confidence-aware rather than unconditional, a point we return to in \Cref{sec:forced}.

\subsection{The active-belief condition is associated with fewer selected low-level errors}
\label{sec:poison}
\begin{figure}[t]
\centering
\includegraphics[width=0.86\linewidth]{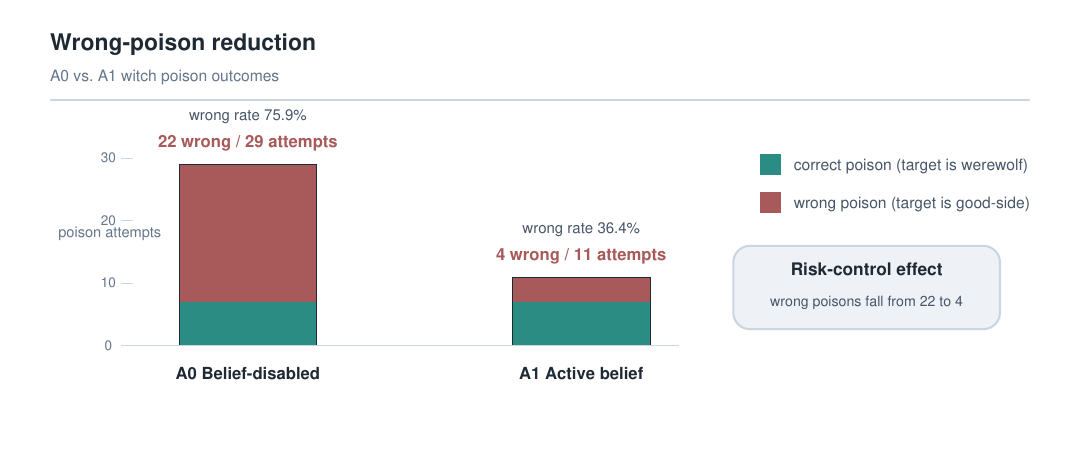}
\caption{Wrong-poison counts, A0 vs.\ A1 ($n=200$ games per arm). The A1 active-belief condition records fewer poison attempts and fewer wrong-poison events than the A0 belief-disabled ablation, with the same number of correct poisons; see Section~\ref{sec:poison} for the Fisher exact test and Wilson intervals.}
\label{fig:poison}
\end{figure}

Beyond the aggregate outcome shift, the active-belief condition also co-occurs with a change in a concrete high-risk action: the witch's poison. Poison is irreversible and, when misdirected, actively helps the werewolves by removing a good-side player. In the paired A0/A1 comparison, this behavior differs on two axes. First, the witch poisons less recklessly: poison is attempted in 29 of 200 games without belief but only 11 of 200 with belief. Second, when poison is used, it is far more accurate: the fraction of poison attempts that hit a good-side player (a wrong poison) drops from 76\% (22 of 29) without belief to 36\% (4 of 11) with belief (\Cref{fig:poison}). The association is significant on a two-sided Fisher exact test ($p=0.03$), but the counts are small: the per-arm Wilson 95\% intervals are wide and partly overlap ($[0.58, 0.88]$ for A0 versus $[0.15, 0.65]$ for A1), so this should be read as directional rather than tightly estimated. We compute the metric by matching \texttt{witch\_poison} events against the post-game role map reconstructed from decision traces, and the resulting wrong-poison count cross-validates with the poison-target-is-wolf count recorded independently in the belief-signal sidecar.

This reduction is consistent with belief acting as a risk-control signal at a concrete, irreversible action: poison is a high-impact, low-frequency decision where a single correct suspicion changes the game, and where even a noisy but usable belief signal can steer the witch away from obviously bad targets. Because poison decisions are infrequent, however, this result should not be read as a complete causal explanation of the aggregate win-rate shift; it is one concrete, auditable behavioral change that co-occurs with the shift.

\subsection{Forced belief consumption fails when belief is flat}
\label{sec:forced}
\Cref{sec:diagnostics} highlighted the low direct action--belief consistency, so a natural intervention is to push agents to align more strongly with belief recommendations. We tested this with a controlled consumption-policy A/B ($n=80$ per arm, seed-paired): a gated baseline versus a variant whose prompt more forcefully requires following the top belief recommendation. The intervention did not deliver a reliable improvement. Action--belief consistency rose only modestly (0.250 to 0.280), and the good-side win rate did not improve (0.412 versus 0.362, a difference well within the $\pm11$pp confidence interval at this sample size). Forcing consumption is therefore not a free lever: it does not reliably translate the available aggregate belief signal into outcomes. \Cref{tab:interventions} collects these intervention results.

\begin{table}[t]
\centering
\caption{Key positive and negative interventions: wrong poison, forced consumption, and load robustness.}
\label{tab:interventions}
\small
\resizebox{\linewidth}{!}{%
\begin{tabular}{lllll}
\toprule
Intervention & Comparison & Metric & Baseline & Treatment \\
\midrule
Belief guidance & A0 vs.\ A1 & poison attempts & 29 / 200 & 11 / 200 \\
Belief guidance & A0 vs.\ A1 & wrong poisons & 22 (76\%) & 4 (36\%) \\
Forced consumption & C1 vs.\ C2 & action--belief consistency & 0.250 & 0.280 \\
Forced consumption & C1 vs.\ C2 & good-side win rate & 0.412 & 0.362 \\
Load robustness & A1 vs.\ L1 & llm\_error rate & 0.00\% & 0.00\% \\
\bottomrule
\end{tabular}}
\end{table}

One plausible explanation is that the stronger consumption rule is effective only when belief is sufficiently confident, but the belief distribution is often flat at key decision points (\Cref{sec:diagnostics}), so there is frequently no sharp recommendation to enforce. When the margin is small, mandating that the LLM follow the top suspect risks turning uncertainty into artificial confidence rather than better decisions.

This supports the design choice in \Cref{sec:method}: belief should be available as a confidence-aware signal, not an unconditional controller. If belief is sharp, deviation deserves scrutiny; if belief is flat, deviation may be harmless or even correct. The remaining headroom identified in \Cref{sec:diagnostics} thus calls for margin-aware consumption, not blanket enforcement.

\subsection{Load confounds can create false strategic conclusions}
\label{sec:load}
The evaluation framework also lets us treat concurrency as an explicit experimental variable rather than an implementation detail. This matters because fallback behavior can be asymmetric across roles: a failed witch action may degrade to a safe skip, while a failed werewolf nomination may be recoverable through other wolves. If LLM error rates rise under load, they can therefore damage one camp more than the other and masquerade as a strategic effect. To keep this from silently contaminating our comparisons, every arm logs its LLM error rate, retry and fallback breakdown, and concurrency setting, and any arm whose error rate departs from the others is analyzed separately rather than treated as a clean strategy comparison.

On our stack this operational confound does not appear in the frozen data. A high-concurrency replica of the active-belief arm (concurrency 32) recorded a 0\% LLM-error rate, matching the low-concurrency arm (concurrency 4), with no context-budget violations, and its good-side win rate stays within the broad range of the other belief arms. Because the load regime is measured rather than assumed, we can state that the main A0/A1 result is not explained by observed load-induced LLM failures, and the same decision traces, fallback statistics, and load metadata would let us detect and quarantine such an artifact if a less robust provider or a higher-load setting reintroduced it.

\subsection{Summary of findings}
The empirical picture is coherent but deliberately guarded. Under paired evaluation, the active-belief condition is associated with a substantially higher good-side win rate (0.205 to 0.390, $p<0.001$), together with usable but not sharp aggregate belief diagnostics (top-1 0.69, top-2 0.83 over belief-targeted sidecar decisions, pooled across camps) and a concrete reduction of irreversible witch-poison errors (wrong poisons 22 to 4). Direct action--belief consistency is low ($\approx0.21$), and camp-restricted arms do not follow a simple holder-benefit pattern, so we do not attribute the outcome shift specifically to belief-content quality; the causal pathway remains unresolved. Attempts to close the consistency gap by forcibly increasing consumption do not help when belief is flat, which argues for margin-aware rather than unconditional consumption policies.

What the results do support is a claim about auditability rather than about intelligence. The same instrumentation that measures the outcome shift also makes it diagnosable: by recording belief updates, contrasting recommendations with actions, and logging load and fallback metadata, the framework lets us surface a concrete error-class reduction, expose that direct consumption is low, reject the tested forced-consumption intervention with evidence, and separate strategy effects from observed infrastructure confounds. External belief is therefore best read as an auditable cognitive baseline that also carries decision-relevant signal, turning opaque agent behavior into replayable, contestable evidence for controlled iteration.

\section{Diagnostic Analyses}
\label{sec:cases}

Aggregate metrics show an outcome shift, but they do not on their own show how the audit framework helps diagnose agent behavior. We therefore present three diagnostic analyses, each illustrating a different use of the framework: surfacing and quantifying a high-impact low-level error class, rejecting an unreliable consumption intervention, and checking whether the win-rate result is entangled with infrastructure artifacts. We note that these are analyses at the arm/aggregate level rather than fully annotated single-game case studies: worked single-game examples with belief curves, event-linked context, and per-decision deviation labels are a natural next artifact of the same instrumentation and are left to future work (Section~\ref{sec:futurework}).

\subsection{Reducing wrong poison actions}
The witch's poison action is one of the highest-risk decisions in the game. It is irreversible, targets a single player, and can directly harm the good side if used on a non-werewolf. In the belief-disabled setting, the witch must infer suspicion from public dialogue, private night information, and role strategy prompts. This creates a common failure mode: the agent may poison based on weak suspicion, local salience, or an overreaction to a single speech or vote.

The belief layer gives the witch an external suspicion baseline. It does not reveal the truth, but it aggregates public evidence into a role-likelihood distribution and exposes top suspected werewolf candidates. When poison decisions are compared against belief and post-game truth, wrong-poison cases become easy to identify: the system can record the target, the belief ranking at the decision point, the public and private evidence visible to the witch, and whether the target was actually on the werewolf side.

In the clean paired comparison, the active-belief condition is associated with better witch behavior on two axes. The witch poisoned less recklessly --- poison was attempted in 29 of 200 games without belief but only 11 of 200 with belief --- and, when used, it was more accurate: the wrong-poison rate fell from 76\% (22 of 29 attempts) to 36\% (4 of 11), reducing absolute wrong poisons from 22 to 4 (Fisher exact $p=0.03$; small, overlapping intervals). This is a concrete, auditable behavioral change on an irreversible action that co-occurs with the aggregate outcome shift in \Cref{sec:winrate}; because poison is infrequent we do not treat it as a full explanation of that shift.

This case also illustrates the benefit of trajectory-level evaluation. By logging belief states, action traces, and post-game truth, the system can link the aggregate outcome shift to a co-occurring change in a specific, auditable error class rather than leaving it to opaque game dynamics. We match \texttt{witch\_poison} events against the post-game role map reconstructed from decision traces to compute this metric, and the resulting wrong-poison count cross-validates with the poison-target-is-wolf count recorded independently in the belief-signal sidecar.

\subsection{Rejecting forced belief consumption}
\Cref{sec:diagnostics} identified a gap between aggregate belief diagnostics and exact action--belief consistency: the diagnostics indicate usable signal, yet agents follow the top recommendation at only about one in five belief-targeted decisions. A straightforward improvement would then be to force agents to align with belief recommendations. We tested this with prompt variants that required stronger belief consumption at relevant decision points.

The intervention did not pay off. In a seed-paired A/B comparison ($n=80$ per arm), the stronger consumption variant raised belief-action consistency only marginally (0.250 to 0.280) and did not improve the good-side win rate (0.412 versus 0.362, within the confidence interval at this sample size). One plausible explanation, supported but not established by the audit data, is that the belief signal is often flat at key decision points, where the top-ranked suspect is not meaningfully more suspicious than the next few candidates. A hard instruction to follow belief is therefore either rarely triggered by confidence gates, or, if applied too aggressively, risks converting weak evidence into false certainty rather than better decisions.

Deviation analysis clarifies the failure. When belief is sharp and the agent deviates, the deviation may indicate a harmful mistake or a strategic override worth inspecting. When belief is flat, deviation from the top belief target is much less informative. Treating all deviations as errors would punish reasonable uncertainty. The failed intervention therefore supports a more conservative policy: belief should be consumed conditionally, and the confidence of the belief state must determine how much authority it receives.

This case is important because it demonstrates that the framework is not merely a tool for justifying belief guidance. It can also reject a proposed improvement. The correct conclusion was not ``make agents obey belief more strongly,'' but ``the current belief signal is often not separable enough to deserve stronger control.'' This distinction is difficult to make without belief snapshots, action traces, and deviation records.

\subsection{Ruling out load-induced confounds}
A third case concerns evaluation validity. A recurring risk in LLM-agent evaluation is that concurrency is treated as an implementation detail: high-concurrency batches can raise LLM timeout and failure rates, and if fallback behavior is asymmetric across roles, those failures can move the outcome and be mistaken for a strategic effect. A failed witch decision can fall back to a skip, weakening the good side's ability to save or poison, while a failed werewolf decision may still be compensated by other werewolves nominating targets. The concern is concrete enough that we treat concurrency as a controlled variable rather than assume it away.

To check whether load introduces an obvious operational confound, we ran a high-concurrency replica (L1) of the active-belief arm at concurrency 32, versus concurrency 4 for the primary arm. The audit metadata shows the two regimes are comparable at the infrastructure level: both record a 0\% LLM-error rate with no context-budget violations, and L1's good-side win rate stays within the broad range of the other belief arms. In other words, on our stack the main A0/A1 effect is not explained by a favorable low-load setting, and no observed asymmetric-fallback confound is present.

This case highlights why agent evaluation in LLM-based environments must report operational metrics alongside strategic outcomes. Because we log LLM error rates, retry counts, fallback breakdowns, and concurrency metadata per arm, we can state that the main result is not explained by observed load-induced LLM failures, and the same instrumentation would let us detect and quarantine a load-induced confound if a less robust provider or a heavier load regime introduced one.

\subsection{Lessons from the diagnostic analyses}
The three analyses illustrate complementary roles of auditability. The wrong-poison analysis shows how belief co-occurs with a reduction in a concrete, high-impact low-level failure, and how the audit layer isolates that specific error class. The forced-consumption case shows how the same framework can reject an intervention that sounds plausible but fails under measured belief flatness. The load case shows how a controlled high-concurrency arm plus operational metrics can rule out an observed LLM-error confound in the frozen data.

Together, these analyses support the paper's central position. In hidden-information LLM-agent systems, the first requirement is not an automatically self-improving agent, but a measurement layer that makes cognition, action, and failure modes inspectable. Once the system can tell whether a failure came from weak evidence, poor consumption, harmful deviation, or infrastructure load, improvement becomes safer and more targeted.

\section{Discussion and Limitations}
\label{sec:discussion}

Our results suggest that external belief plays a dual role in hidden-information LLM-agent systems. In our setting, the active-belief condition is associated with a higher good-side win rate under paired evaluation, so belief carries decision-relevant signal rather than being merely a visualization. We stop short, however, of calling this a demonstrated improvement in reasoning: the low direct consistency, the camp-restricted arms, and the belief-content ablation baseline leave open whether the effect comes from belief content, prompt structure, or indirect behavioral change. Belief is therefore best treated primarily as an auditable baseline for diagnosis, intervention testing, and safer iteration --- useful as decision support where it is confident, and as a measurement layer everywhere else.

\subsection{What belief is useful for}
The strongest value of the belief layer is observability. Without belief, an LLM agent's action can be logged, but it is difficult to know what hidden-state inference the action implied. With belief, the system can compare an action against an explicit cognitive baseline. If the agent votes away from the top suspected werewolf, the deviation can be inspected. If the witch poisons a player with low werewolf probability, the case can be flagged. If belief is flat, the system can avoid over-interpreting disagreement.

Belief is also useful as a diagnostic compression of history. A Werewolf game contains many speeches, votes, claims, deaths, and role-specific observations. Feeding the full history into every prompt is costly and does not guarantee reliable reasoning. A belief snapshot provides a compact, replayable summary of how the system currently interprets the hidden state. Because belief updates are tied to structured events, the summary can be inspected after the game rather than treated as an opaque memory.

Finally, belief supports safer improvement. In high-noise games, changing prompts or rules based on final win rate alone is risky. Belief-action deviations provide more local evidence: which decision contradicted the current suspicion baseline, whether the baseline was sharp enough to trust, and whether the deviation helped or harmed after truth was revealed. This makes strategy updates more targeted than direct win-loss feedback.

\subsection{What belief cannot solve}
Belief is not a substitute for evidence. If public information does not separate werewolves from good-side players, then a belief model should remain uncertain. In our experiments, good-side belief was often flat at key decision points. This is not necessarily a failure of the update rule; it can also be an honest representation of limited public evidence. A belief system that becomes overconfident without strong evidence may look more decisive, but would be less reliable.

Belief is also not a complete policy. It can suggest which players look suspicious, but it does not determine how an agent should speak, when a seer should reveal, how a werewolf should bluff, or whether a deviation is strategically useful. These behaviors still depend on role strategy, natural-language context, and long-horizon social reasoning. The observed outcome shift is partial, which is consistent with belief informing selected decisions without specifying the full role policy.

Nor should belief be consumed unconditionally. Our forced-consumption experiments show that stronger belief-following instructions can fail when the belief signal is weak. A flat belief distribution does not become more useful because the prompt tells the agent to obey it. Consumption policy must therefore depend on confidence, separability, and decision context.

\subsection{Implications for hidden-information agent evaluation}
The main implication is that hidden-information LLM agents require trajectory-level evaluation. Final win rate is too coarse to distinguish between weak inference, poor role strategy, harmful deviation, reasonable strategic override, and infrastructure failure. A large number of games can still support a wrong conclusion if error rates, fallback behavior, model load, or information leakage are not tracked.

Our load-confound case illustrates this point. By treating high concurrency as a separate arm rather than silently mixing it into the main strategy comparison, we can check whether the frozen results are driven by load-induced LLM failures or asymmetric fallbacks. This kind of operational confound is likely to appear in other LLM-agent environments as well, especially when agents depend on external model APIs, long prompts, or multi-turn orchestration. For this reason, evaluation infrastructure should be treated as part of the method, not merely as engineering support. Information isolation, structured logging, replay, decision traces, belief snapshots, and fallback accounting are all necessary for drawing credible conclusions about agent behavior under hidden information.

\subsection{Generalization beyond Werewolf}
Werewolf is a convenient testbed, but the underlying structure appears in many real-world settings. Fraud detection, negotiation, cybersecurity, supply-chain risk assessment, and geopolitical forecasting all involve hidden state, strategic behavior, noisy feedback, and a need for auditability. In such settings, it may be unsafe to let a black-box LLM update its behavior directly from single outcomes. The framework studied here suggests a more conservative pattern. Maintain an external, inspectable state estimate; compare model decisions against that estimate; diagnose deviations; aggregate evidence across cases; and update strategy only through reviewed, versioned changes. The specific belief features used for Werewolf will not transfer directly, but the separation between cognition baseline, action generation, deviation analysis, and offline improvement is more general.

\subsection{Limitations}
This work has several limitations. First, the environment is still a game. Although Werewolf captures hidden roles, deception, partial observability, and high-variance feedback, it is simpler and more controlled than real-world adversarial domains. Claims about transfer should therefore be understood as architectural hypotheses, not as direct empirical validation.

Second, the belief rules are partly human-designed. This makes them interpretable and easy to audit, but also limits coverage. The current system may miss subtle linguistic signals, long-horizon deception patterns, or role-specific tactics that experienced human players would recognize. Future work could learn parts of the belief update function from data, provided that learned updates remain inspectable and are evaluated against leakage risks.

Third, the experiments depend on model quality, prompt versions, and infrastructure conditions. Different LLMs may consume belief differently, produce different levels of strategic deception, or fail under different load conditions. This is why we report operational metrics such as LLM errors and fallback rates, but broader cross-model evaluation remains necessary.

Fourth, our framework does not implement full self-evolution. The offline loop produces candidate insights and supports reviewed strategy updates, but it does not automatically rewrite prompts, update belief rules, or fine-tune model weights without human oversight. This is a deliberate safety choice in a high-noise environment, but it means the system is not a complete autonomous learning agent.

Fifth, the current belief layer is not a full theory-of-mind model. It tracks role likelihoods and selected suspicion relationships, but does not deeply model nested beliefs such as ``what player A believes player B believes about player C.'' Such recursive reasoning may be important for more advanced deception and persuasion, but it also introduces additional complexity and audit challenges.

Sixth, and most importantly for interpretation, the mechanism behind the win-rate gain is not fully identified. Giving belief only to the werewolves raises the good-side win rate more (wolves-only, 0.313) than giving it only to the good side (villagers-only, 0.253), which is not what a simple ``belief improves the holder's hidden-role inference'' account would predict. Combined with the low action--belief consistency ($\approx0.21$), this suggests that part of the effect may operate through changes in agent behavior induced by the presence of the belief section rather than through the belief content being consumed at the decision. Two controls would sharpen the attribution and are left to future work: a randomized or shuffled-belief arm that preserves the prompt structure while destroying belief content, and a natively designed pure-LLM prompt as the baseline (rather than the belief-oriented prompt with belief removed, as in A0). Until these controls are run, the win-rate result should be read as an effect of the active-belief \emph{condition} rather than as established evidence that belief content improves inference.

\subsection{Future work}
\label{sec:futurework}
The most important next step is to isolate the mechanism behind the outcome shift. Two controlled arms would separate belief content from prompt structure: a native no-belief baseline (\emph{A0-native}) built from a prompt that never contained a belief section, to test whether A0's belief-content ablation handicaps the baseline; and a randomized-belief arm (\emph{A-rand}) that preserves the prompt structure but shuffles or permutes the belief content, to test whether the effect is driven by belief information or by the presence and formatting of the belief section. A small-scale replication of A0-native, A1, and A-rand on a second model would indicate whether the direction of the effect is model-specific. In parallel, recomputing camp-separated, vote-time diagnostics (good-side-only top-1/top-2 with dynamic per-decision chance baselines) and quantifying deviations (frequency, the harmful/justified/strategic distribution, an annotation protocol with inter-rater agreement, and belief-on/off comparisons) would turn the current instrumentation and conceptual taxonomy into validated measurements. Fully annotated single-game trajectories---belief curves, event-linked context, and per-decision deviation labels---would complement these aggregate analyses.

A further direction is improving the quality of the belief signal itself. Separability at key decision points is moderate, so better extraction of public claims, vote rationales, contradiction patterns, and seer-claim dynamics may provide sharper evidence than simply increasing belief weights.

A second direction is role-specific belief. Werewolves already know who their teammates are, so a generic ``who is a werewolf'' belief is less useful to them. More useful wolf-side belief may estimate who is likely to be the seer, witch, or hunter. Because the system already represents role distributions, this extension can be studied without changing the basic belief abstraction.

A third direction is broader controlled evaluation. Future experiments should compare multiple LLM providers, larger seed ranges, different concurrency regimes, and human-in-the-loop mixed games. Such studies would clarify which findings are robust properties of hidden-information agents and which are artifacts of a specific model or deployment condition.

Finally, the offline improvement loop can be extended toward more automated learning, but only after the evidence pipeline is stable. High-quality trajectories with belief snapshots, action traces, deviation labels, and reviewed strategy outcomes could become training data for prompt learning, retrieval policies, or fine-tuning. The key requirement is that learning should inherit the auditability of the data pipeline rather than collapse back into an opaque black-box update.

\section{Conclusion}
\label{sec:conclusion}

We studied LLM agents in a 9-player Werewolf environment as a testbed for hidden-information multi-agent decision making. We introduced an auditable framework that combines strict information isolation, structured event and decision traces, an external belief layer, deviation-based diagnosis, and a defensive offline improvement loop.

Our 1,080-game frozen evaluation, including a 200-seed paired A/B comparison, shows that the active-belief condition is associated with a significantly higher good-side win rate (0.205 to 0.390, $p<0.001$) and with fewer irreversible low-level errors such as wrong witch poisons, while aggregate belief diagnostics indicate usable hidden-role signal. Crucially, it also shows what these experiments do \emph{not} establish: direct action--belief consistency is low, giving belief only to the werewolves helps the good side more than giving it only to the good side, and the belief-disabled baseline retains the belief-oriented prompt structure, so the outcome shift cannot yet be attributed specifically to belief content. What the framework does deliver is auditability: the same instrumentation that measures the shift also exposes low direct action--belief consistency, rejects an unreliable forced-consumption intervention with evidence, and distinguishes strategy effects from observed infrastructure confounds such as high-concurrency load. Belief thus serves less as a proven controller of play and more as an auditable layer that turns opaque agent behavior into replayable, contestable evidence.

These results argue for treating belief as an auditable cognitive baseline rather than a direct controller. In high-noise partially observable settings, safer improvement requires first making agent reasoning measurable and replayable. Werewolf is one controlled testbed, but the same principle applies more broadly to LLM agents operating under hidden state, strategic behavior, and noisy feedback.

\section*{Use of Generative AI Tools}
Generative AI language tools were used to assist with language editing, structural revision, and code-level manuscript review. All research questions, system design, experiments, analyses, citations, and final claims were reviewed and verified by the human authors, who take full responsibility for the content.

\section*{Code and Data Availability}
The project repository is available at \url{https://github.com/JJJayden-Yang/ai-werewolf}. It contains the implementation and will host the paper-facing release artifacts. The frozen 1{,}080-game experimental corpus used in this paper, including events, decision traces, belief snapshots, per-arm reports, and run provenance, is being prepared as an archived release linked from the repository to support replay and re-computation of the reported metrics.


\bibliographystyle{plainnat}
\bibliography{references}

\end{document}